# Seismological Aspects of December 31, 2018, Cairo-Suez Event

*Mohamed N. ElGabry\*, Hesham M. Hussein, Mona Abdelazim, Adel S. Othman, Shimaa Hosny, and Hany M. Hassan*

*Egyptian National Data Center*

*National Research Institute of Astronomy and Geophysics, NRIAG, 11421 Helwan, Egypt*

*\*elgabry@nriag.sci.eg*

## Introduction

On 31 December 2018 at 10:36:34 UTC shallow earthquake of $M_L$=4.3 took place 25 Km east of Cairo and in the vicinity of the under-construction new Capital city of Egypt (Fig.1). This earthquake is located using the HYPOINVERS software. Some 21 stations were used for estimating the location of this event including eight s-phases and 13 p-phases. The horizontal error is about 1.2 Km, while the depth error is 0.8 Km. The estimated maximum azimuth gap is 50º. The earthquake was widely felt within an area of about 55km (Cairo, New Capital, and Helwan) with a maximum experienced intensity of IV on the European Macroseismic Scale (EMS). The earthquake was felt indoors by many people, outdoors by few as reported by social media, local newspapers and field questionnaire. The level of vibration caused by this comparably small size earthquake was frightening for many people. Some have reported swing of hanging objects at some places, and no damage to buildings was observed.

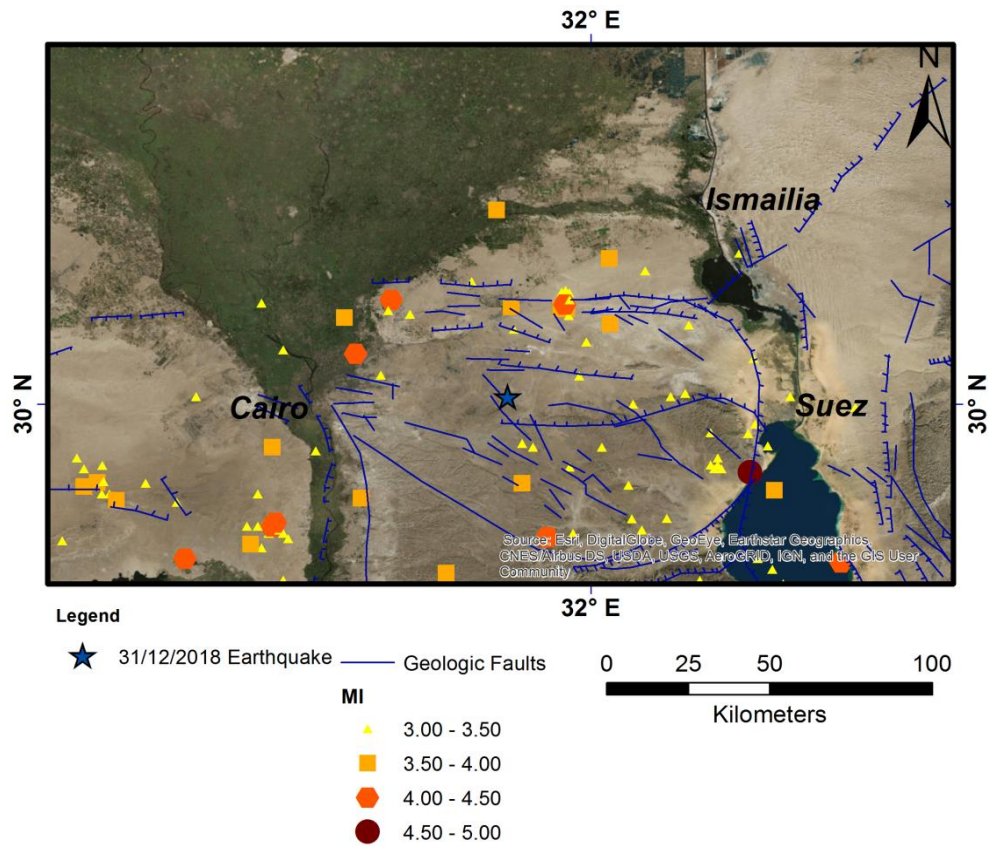

*Fig. 1: Seismicity of the Cairo-Suez district area for the period from 1997-2018 for Ml≥3 plotted with the geologic faults from EGSMA, (1981). The blue star is the location of the 31 Dec 2018 event.*

**Seismotectonics of Cairo-Suez District CSD**

The event occurred in CSD, which is considered as one of the well known active seismogenic zones in Egypt, and it is located in the northern part of Eastern Desert of Egypt and covers the area that extends from the northern end of the Suez rift to the Nile Valley. This region represents a source of a potential seismic threat as evidenced by GPS campaigns (~2mm/year by Mahmoud et al., 2005 and ~5mm/year by Badawy, 2001), the occurrence of moderate size earthquakes and the presence of morphostructural indicators (e.g., fault scarps) (Gorshkov et al., 2018).

The CSD is affected by the late Oligocene-early Miocene deformation related to the opening of the Gulf of Suez in response to ENE-WSW oriented extension. Meanwhile, it is probable that part of this deformation is transferred to the land and led to the rejuvenation of the deep-seated preexisting E-W oriented faults by dextral transition (oblique-slip movement) in addition to NW-SE striking faults (Moustafa and Abd Allah, 1992). This movement had generated E-W elongated belts of left stepped en-echelon normal faults (Moustafa et al., 1985), overlie preexisting deep-seated faults of right lateral strike-slip movement (Smith, 1965; Moustafa, 1988). Those belts are consistently found throughout the area and act as transfer zones between the NW oriented normal faults (i.e., to transfer the throw from one NW oriented fault to another) (Fig. 1).

Although the CSD is characterized by moderate to low seismicity the spatial analysis of seismicity can reveal the clusterization of seismicity in three distinct clusters: Wadi Hagul, Abu-hammed, and bitter lakes, respectively. Most of the clustered seismicity is well consistent with the main E-W trend, in particular to the south while in the northern part the seismicity is diffused and it is difficult to attribute it to any of the known geologic structure. The majority of this activity is more likely conformable with preexisting NW-SE, E-W, and NNW dominant surface faults. The focal mechanism solutions for events in this region are predominantly dip-slip normal fault trending in NW-SE and the E-W directions with slight strike-slip component, as shown in Fig. 2. Hussein et al. (2013) have calculated the minimum principal stress to be oriented $N30 E^0$, which supports the model of the rejuvenation of the Pre-Tertiary E-W faults as a continuation of the Gulf of Suez extensional process on land.

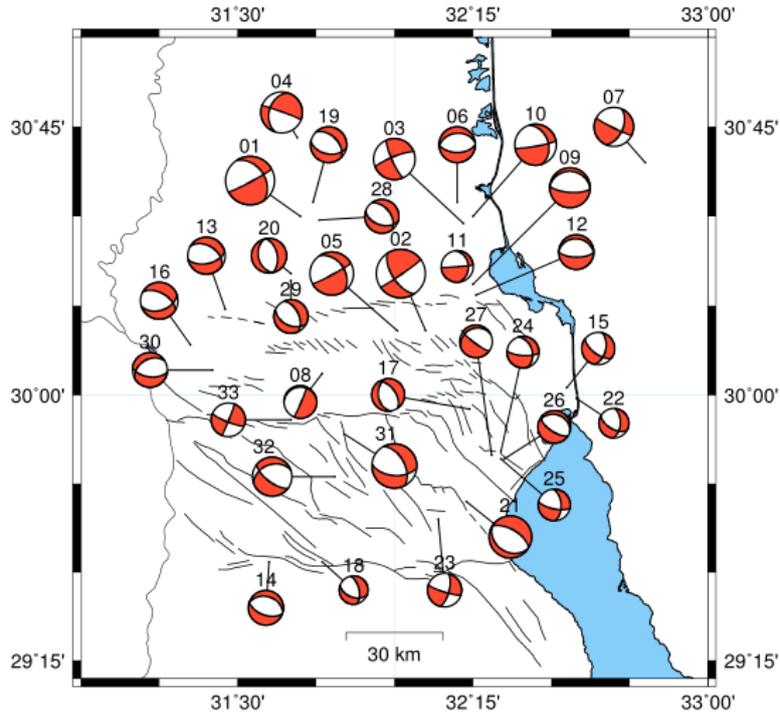

*Fig. 2: Focal mechanism solutions of the events occurred in the study area (for more information see Appendix).*

**Source parameters and mechanism**

The focal mechanism solution of the Dec. 31 2018 earthquake wasconstructed using data from 31 stations in Egyptian National Seismic Network (ENSN), Egyptian Strong Motion Accelerograph Network (ESMA) and International Monitoring System (IMS). The initial focal mechanism solution was constructed manually from the polarity of the first onset of the P-wave using PMAN program of Suetsugu (1998). The final solution was constructed by using FOCMEC software (Snoke et al., 1984) by the polarity of first onset P-wave, and $S_H$, $S_V$ polarities, in addition to the spectral amplitude ratios for $S_V/P$ and $S_H/P$ these additional information make solution more reliable solution. The obtained focal mechanism solution shows normal faulting mechanism with a minor component of strike-slip with two nodal planes trending almost NNW-SSE and WWN-WWS-trending fault planes (Fig. 3), which is consistent with the tectonic regime of the CSD. Table 1 shows the parameters of the fault plane solutions.

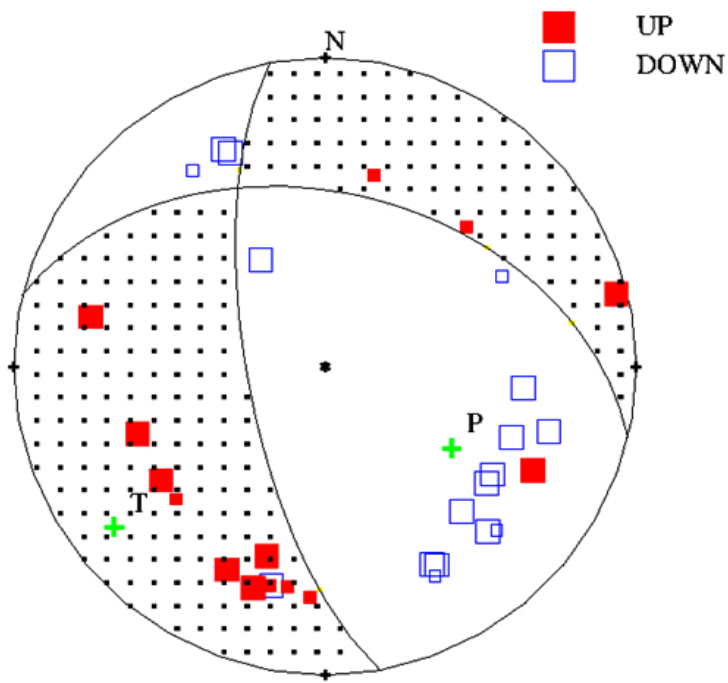

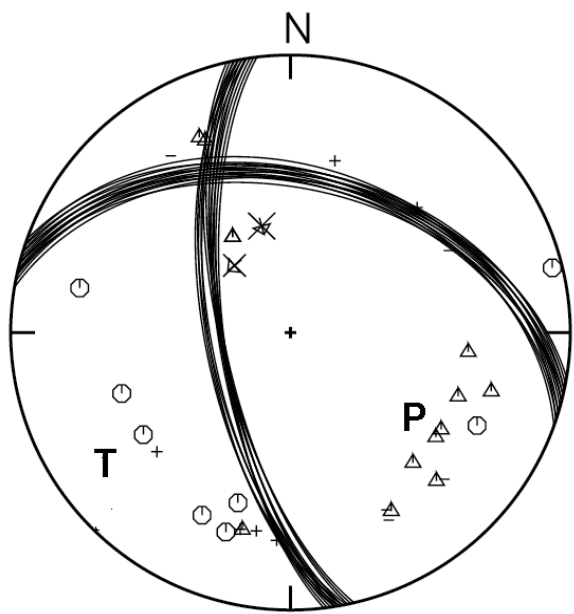

Fig. 3: The focal mechanism solution of the 31 Dec 2018 Earthquake; up is first polarity solution using PMAN software of Suetsugu (1998); down is the accepted solution constructed joint polarity and amplitude ratio solution using FOCMEC software of Snoke et al., (1984).

Table 1. Parameters of the accepted fault plane solution.

| Strike | Dip | Slip | P- axis | T-axis |
|--------|-----|------|---------|--------|
|        |     |      |         |        |

| | | | Azimuth | Plunge | Azimuth | Plunge |
|---|---|---|---|---|---|---|
| 283 | 43 | -149 | 123 | 50 | 52 | 16 |

To provide insight analysis for this event, the seismic source parameter is another useful information to evaluate the ground motion for seismic hazard assessment. The seismic moment (M0), corner frequency (fc), fault radius (r), stress drop (Δσ) and the moment magnitude (Mw) was determined for the recorded stations that have a good signal to noise ratio using SH-wave amplitude displacement spectra. The analyzed stations are located at different epicentral distances ranging from 11 to 130 km. To estimate the spectral amplitude and source parameters of the earthquake, the EQK_SRC_PARA code was applied, (Kumar et al., 2012). The SH spectrum is calculated from rotated horizontal broadband seismic records which are transformed into frequency domain using Fast Fourier Transform (FFT) and corrected for attenuation based on a frequency dependent attenuation correction relation $Q_C = 85.68 f^{0.79}$ (EL-Hadidy et al., 2006). The Brune's source spectral model (Brune, 1970 and 1971), was applied using a non-linear inversion Leven-berg Marquardt technique (Press et al., 1992). The spectral analysis and the source parameters obtained for the event are listed in Table 2 whereas an the fitted displacement spectra are plotted in Appendex Fig. A,B,C. The average source parameters are calulated using the equations of Archuleta et al. (1982).

Table 2: Source parameters of 31 December 2018 earthquake from HAG, ZAF and NAT stations.

| Station code | Δ (Km) | $AZ°$ | $f_c$(HZ) | $M_0\ (Nm)$ | r (Km) | Δσ (MPa) | $M_w$ |
|---|---|---|---|---|---|---|---|
| HAG | 33 | 102 | 5.1 | 3.9E+14 | 0.261 | 9.55 | 3.7 |
| ZAF | 111 | 137 | 5.1 | 5.8E+14 | 0.261 | 14.38 | 3.8 |
| NAT | 120 | 250 | 4.8 | 8.8E+14 | 0.278 | 17.99 | 3.9 |
| Average Value | | | 5 | 5.83E+14 | 0.269 | 13.49 | 3.8 |
| SD | | | ±0.12 | ±0.18 | ±0.02 | ±0.14 | ±0.03 |

**Conclusions**

The source parameters of the December 31 2018 new Capital city earthquake which was detected by the stations of the Egyptian National Seismic Network (ENSN), Egyptian Strong Motion Accelerograph Network (ESMA) and International Monitoring System (IMS) have been estimated from the displacement spectrum. Three stations at different epicentral distances were investigated to find these parameters. For this earthquake, we obtained a value of 0.261 km for the fault radius, 13.49 MPa for the stress drop and 3.8 for the moment magnitude Mw. The stress drops of this event is located within those range of intraplate earthquakes explained by Allmann and shearer (2009).

The focal mechanism solution was also constructed from both first polarity and amplitude ratio. It exhibits normal faulting mechanism with a minor component of strike-slip. The minimum principal stress is found to be N50 $E^0$ which in good correlation with N30 $E^0$ value calculated from the stress field inversion by Hussein et al. (2013) and neotectonics studies by Moustafa and Abd Allah, (1992). These results could support that the event is triggered by the Pre-Tertiary E-W faults as an on land continuation of the Gulf of Suez extensional process.

This event was felt to distances up to 50 Km away from the epicenter. The Maximum observed Macroseismic intensity is IV on EMS.

**Data and resources**

The waveform data were provided by Egyptian National Seismological Network ENSN, Egyptian Strong Motion Accelerograph network ESMA, International Monitoring System IMS.

# Appendix

Table 1: Focal mechanisms solutions for the CSD area used in Figure 2.

| STRIKE | DIP | RAKE | MAG | INDEX |
| --- | --- | --- | --- | --- |
| 326 | 40 | -7 | 4.6 | 1 |
| 147 | 62 | 4 | 4.6 | 2 |
| 248 | 80 | -170 | 3.9 | 3 |
| 197 | 40 | -4 | 3.9 | 4 |
| 326 | 40 | -7 | 4.1 | 5 |
| 263 | 36 | -96 | 3.4 | 6 |
| 23 | 52 | -6 | 3.7 | 7 |
| 267 | 9 | -26 | 3.1 | 8 |
| 94 | 71 | -90 | 3.8 | 9 |
| 342 | 38 | -10 | 3.9 | 10 |
| 353 | 30 | -3 | 3 | 11 |
| 275 | 24 | -86 | 3.3 | 12 |
| 302 | 45 | -56 | 3.5 | 13 |
| 286 | 35 | -93 | 3.3 | 14 |
| 13 | 52 | -36 | 3.1 | 15 |
| 72 | 46 | -127 | 3.5 | 16 |
| 172 | 40 | -69 | 3.1 | 17 |
| 110 | 48 | -136 | 2.7 | 18 |
| 315 | 48 | -66 | 3.4 | 19 |
| 167 | 39 | -91 | 3.3 | 20 |
| 123 | 29 | -88 | 4 | 21 |
| 3 | 54 | -41 | 2.8 | 22 |
| 14 | 68 | -10 | 3.2 | 23 |
| 94 | 72 | -128 | 3 | 24 |
| 102 | 70 | -142 | 3 | 25 |
| 124 | 66 | -83 | 3.1 | 26 |
| 243 | 21 | -152 | 3.1 | 27 |
| 323 | 48 | -65 | 3.2 | 28 |
| 116 | 45 | -120 | 3.2 | 29 |
| 97 | 57 | -67 | 3.3 | 30 |
| 332 | 53 | -47 | 4.2 | 31 |
| 250 | 38 | -141 | 3.7 | 32 |
| 112 | 79 | -177 | 3.2 | 33 |

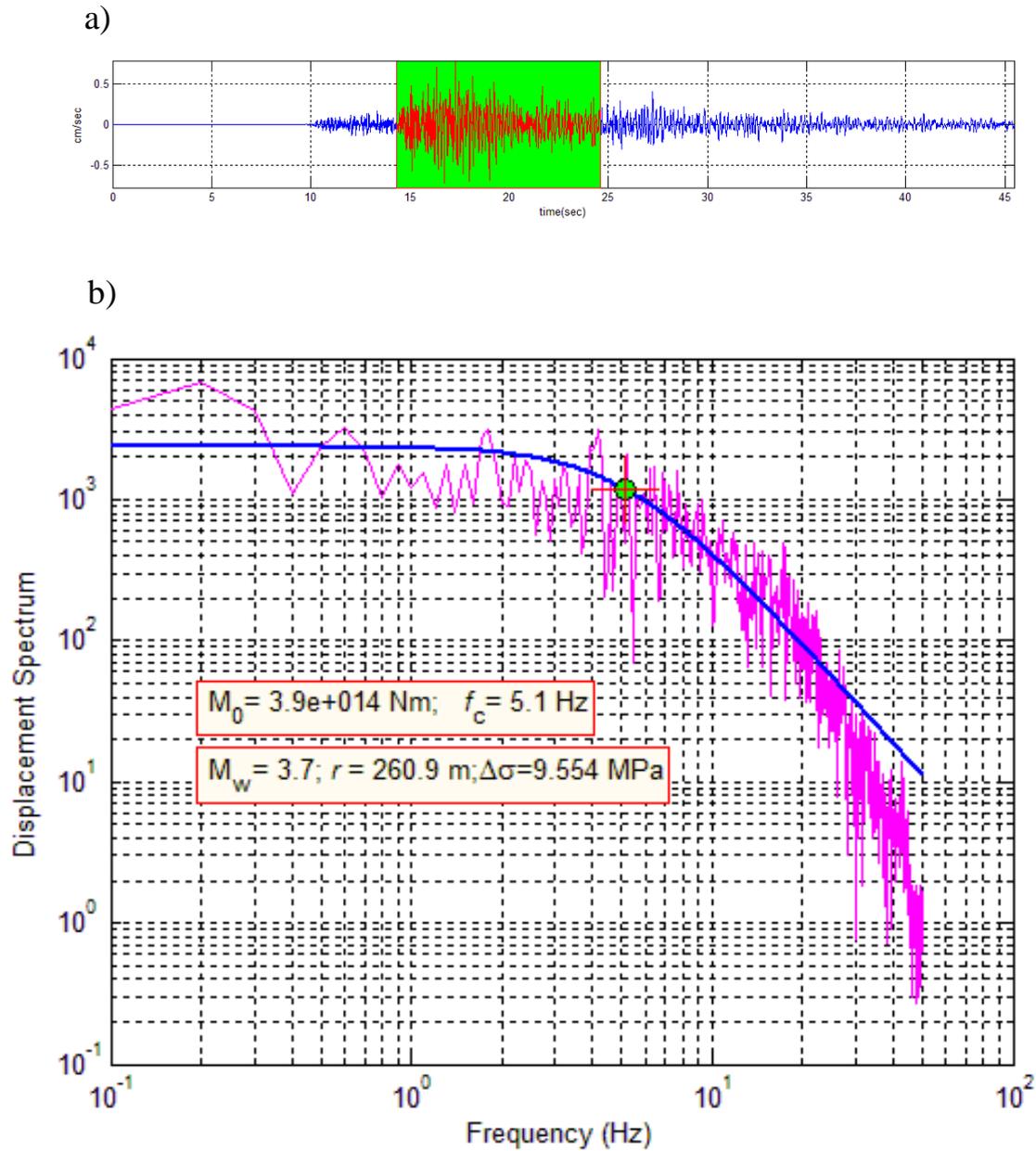

Fig. A: a) Recorded velocity seismogram of HAG seismic station that located at 33 km from the epicenter of the earthquake and b) Corrected displacement spectra and source parameters values.

a)

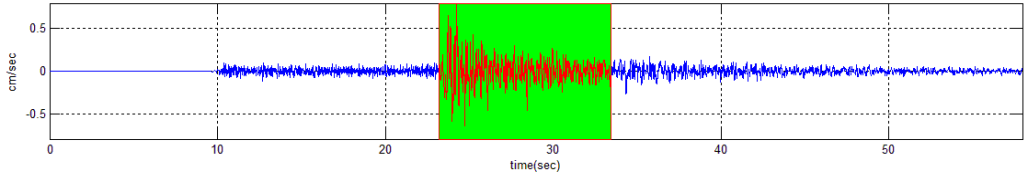

b)

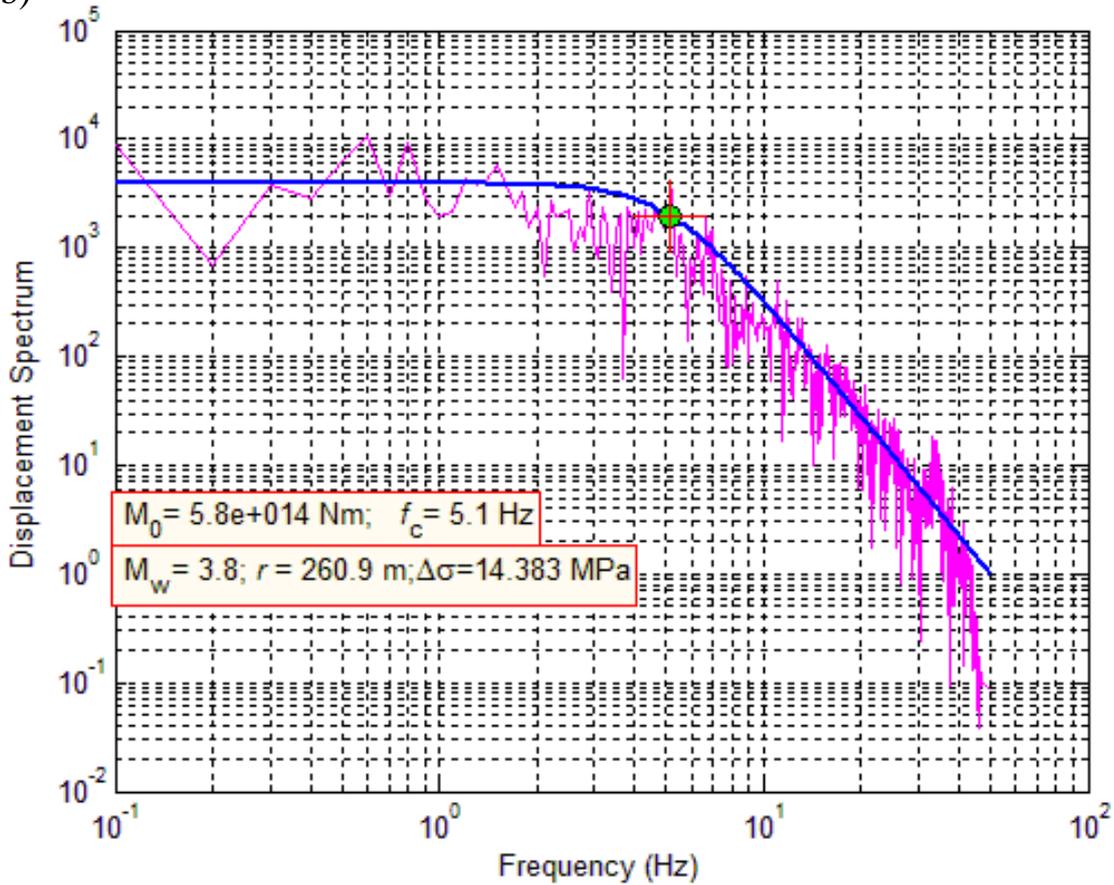

Fig. B: a) Recorded velocity seismogram of ZAF seismic station that located at 111 km from the epicenter of the earthquake and b) Corrected displacement spectra and source parameters values.

a)

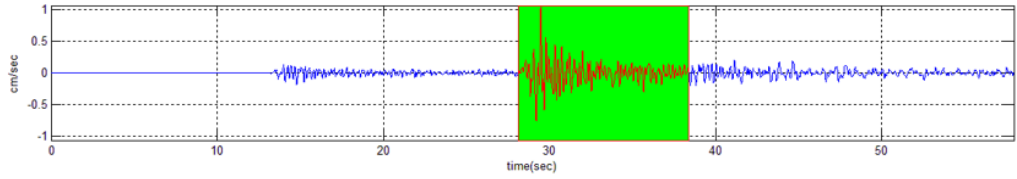

b)

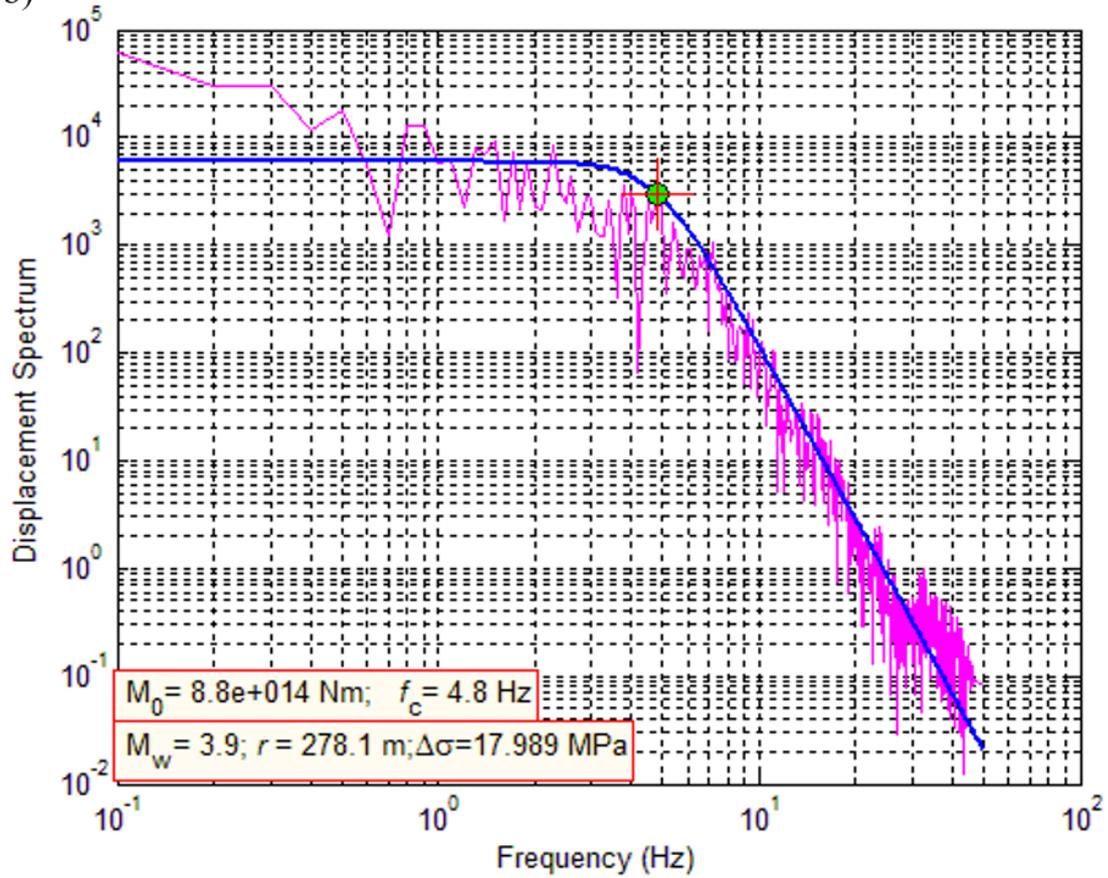

Fig. c: a) Recorded velocity seismogram of NAT seismic station that located at 120 km from the epicenter of the earthquake and b) Corrected displacement spectra and source parameters values.